\documentclass[oneside,a4paper,reqno]{amsart}
\usepackage{amsmath}
\usepackage{amssymb}
\usepackage{eucal}
\usepackage{graphics}

\newcommand{\bmat}[3]{\bigl \langle  
       #1 \bigr| \, #2\, \bigl|     #3  
                  \bigr \rangle}

\newcommand{\ket}[1]{		\left| #1 \right>  }

\newcommand{\bra}[1]{		\left< #1 \right|  }

\begin{document}
\begin{titlepage}
\begin{center}
\bfseries CONTEXTUALITY OF APPROXIMATE MEASUREMENTS
\end{center}
\vspace{1 cm}
\begin{center} D M APPLEBY
\end{center}
\begin{center} Department of Physics, Queen Mary
and
		Westfield College,  Mile End Rd, London E1 4NS,
UK
 \end{center}
\vspace{0.5 cm}
\begin{center}
  (E-mail:  D.M.Appleby@qmw.ac.uk)
\end{center}
\vspace{0.75 cm}
\vspace{1.25 cm}
\begin{center}
\textbf{Abstract}\\
\vspace{0.35 cm}
\parbox{10.5 cm }{ The claim of Meyer,
Kent and Clifton (MKC) that finite precision
measurement nullifies the Kochen-Specker theorem is
criticised.  It is argued that, although MKC
have nullified the Kochen-Specker theorem strictly
so-called, there are other, related propositions which
are not nullified.  The argument given is an
elaboration of some of Mermin's critical remarks.
Although MKC allow for the fact that the observables
to be measured cannot be precisely specified, they
continue to assume that the observables which are
actually measured are strictly commuting.  As Mermin
points out, this assumption is unjustified. 
Consequently, the analysis of MKC is incomplete.  To
make it complete one needs to investigate the
predictions their models make regarding approximate
joint measurements of non-commuting observables.  
Such an investigation is carried out, using methods
previously developed in connection with approximate
joint measurements of position and momentum.  It is
shown that a form of contextuality then re-emerges.
                      }
\end{center}
\vspace{1 cm}
\begin{center}
\end{center}
\end{titlepage}
\section{Introduction}
\label{sec:  intro}
In a recent series of papers Meyer~\cite{Meyer},
Kent~\cite{KentA} and
Clifton and Kent~\cite{KentB} (to whom we will
subsequently refer
  as MKC) claim to have ``nullified'' the
Kochen-Specker
theorem~\cite{Koch,Bell,MerminB,Peres}.  They infer
that ``there is no truly compelling argument
establishing that non-relativistic quantum mechanics
describes classically inexplicable
physics''~\cite{KentB}.  They suggest that this may
have significant consequences for quantum information
theory and quantum computing.

The purpose of this paper is
to criticize MKC's conclusions.  
It is true that MKC
have circumvented the
\emph{particular kind} of non-classicality which
features in the Kochen-Specker theorem.  It is also
true that in doing so they have significantly
deepened our understanding of the conceptual
implications of quantum mechanics.   However, when it
comes to the central question, as to whether
non-relativistic quantum mechanics is classically
explicable, it appears to us that a closer examination
of their models leads to a different conclusion. 
We will argue that, although MKC have nullified the
Kochen-Specker theorem \emph{strictly so-called},
there are other,  related propositions which are
not nullified. The argument we will give is a
development of some of the points made in Mermin's
critique~\cite{Mermin} (for other critical comments see
Havlicek \emph{et al}~\cite{Havli}, 
Cabello~\cite{Cabel} and Basu~\emph{et
al}~\cite{Basu}).  

As
Mermin points out, MKC's analysis of finite precision
measurements is not entirely adequate.  MKC only
consider  one source of non-ideality:  namely, the
non-ideality which is due to inaccuracies in the
specification of the observables to be measured.  In
every other respect the measurements they 
consider\footnote{In 
the main part of their argument.
 The part of their argument which concerns (in their
terminology) ``positive operator measurements'' will
be discussed below (see Section~\ref{sec: 
CKonPOVMs}).
} 
are perfectly ideal (using the word ``ideal'' in the
sense defined in Section~\ref{sec:  MeasThy}).  Such
measurements might be rather better described
(following Mermin) as ideal measurements which are not
precisely specified.  Consequently, the analysis of
MKC is incomplete.  In order to make it complete one
needs to extend the analysis to the case of
measurements which are not ideal \emph{in any
respect}:  not ideal in respect of the target
observables, which the apparatus is intended to
measure; and not ideal in respect of any other
observables either.  The purpose of this paper is to
present such an extended analysis.  In the first part
of the paper we give a more comprehensive account of
approximate quantum mechanical measurements (based on
ideas previously presented in
Appleby~\cite{self1a,self1b}).  In the second part we
apply  these results to the MKC models.

It is important to distinguish the specific, technical
result proved by Kochen and Specker, and the
essential point of their argument.  By the ``essential
point''  we mean the
proposition that quantum mechanics (whether
relativistic or not) is inconsistent with  classical
conceptions of physical reality. 

In the theories of
classical physics it was  tacitly assumed 
\begin{enumerate}
\item To each observable quantity characterising
a   system there corresponds an objective
physical quantity, which has a determinate value
at every instant.
\label{en:  AOVObj}
\item An ideal, perfectly precise measurement
gives, \emph{with certainty}, a value which 
\emph{exactly
coincides} with the value which the quantity being
measured objectively did possess, immediately
before the measurement process was initiated.
\label{en:  AOVIdeal}
\end{enumerate}
Of course, real laboratory measurements   
are not perfectly precise; and this fact 
was acknowledged in classical physics, just as it
is in quantum physics.   Consequently, the above
propositions ought to be supplemented:
\begin{enumerate}
\setcounter{enumi}{2}
\item A non-ideal, approximate measurement gives, 
\emph{with high
probability}, a value which is \emph{close} 
to the value
which the quantity being measured objectively did
possess, immediately before the measurement
process was initiated.
\label{en:  AOVProx}
\end{enumerate}
We will refer to
these three propositions collectively as the
principle of accessible objective values, or the
AOV principle for short.  Of course, if the
AOV principle  is not true, it does not necessarily
follow that objective values do not exist.  However,
if the postulated objective values are typically quite
different from the values obtained by measurement,
then it is difficult  to see what is achieved by
assuming them.  Consequently, failure of the AOV
principle can be taken (though need not necessarily be
taken) to justify a positivistic view:  on the
grounds  that ``a wheel that can be turned though
nothing else moves with it, is not part of the
mechanism'' (as Wittgenstein~\cite{Witt} succinctly
put it, in a  different context).  This was (in
essence) the perception which motivated the Copenhagen
Interpretation.

 The
significance of the Kochen-Specker theorem is that it
seems to provide a  rigorous proof that
the AOV principle is inconsistent with the
predictions of quantum mechanics.  
Kochen and Specker show that, if it is possible 
to make joint, ideal measurements of any set of
commuting observables then, in a hidden variables
theory, the result of making an ideal measurement of
one observable must, in general, depend on which other
commuting observables are jointly and ideally measured
with it.  This property is not consistent with 
clause~\ref{en:  AOVIdeal}  of the AOV principle. 

The weakness in Kochen and Specker's argument was
identified by MKC, who noted that  the observables to
be measured cannot be specified with perfect
precision.  Consequently, in an experiment which is
intended to measure one set of commuting observables,
the possibility cannot be excluded  that what is
actually measured is another, slightly different set of
commuting observables.  MKC use this freedom to
construct a hidden variables theory which
\emph{does} satisfy clause~\ref{en:  AOVIdeal} of the
AOV principle.  It should be noted that the theory
they construct is not strictly equivalent to standard
quantum mechanics (because they  postulate that an
observable can only be measured if it belongs to a
particular, proper subset of the set of all
self-adjoint operators).  Consequently, they have not
shown that clause~\ref{en:  AOVIdeal} of the AOV
principle is consistent with  all the  predictions of
quantum mechanics (\emph{i.e.},  they have not
\emph{refuted} the Kochen-Specker theorem). 
On the other hand, they have  shown that this clause
is consistent with the predictions of quantum mechanics
in so far as these are empirically verifiable
[\emph{i.e.}, they have  \emph{nullified}
the Kochen-Specker theorem (strictly so-called)].

However, just as MKC have noted a significant weakness
in the argument of Kochen and Specker, so in turn
Mermin~\cite{Mermin} has noted a significant weakness
in theirs.  Although MKC allow for the fact that the
observables which are actually measured may be
slightly and uncontrollably different from the target
observables, which the experiment is intended to
measure, they nevertheless follow Kochen and Specker in
assuming that the observables which are actually
measured still are strictly commuting.  But, as Mermin
points out, MKC's own assumptions suggest that the
observables which are actually measured will almost
certainly \emph{not} be strictly commuting.  Also
(and, as it turns out, closely connected with this
point) one may ask:  do the MKC models satisfy 
clause~\ref{en:  AOVProx} of the AOV principle?  After
all, if one accepts MKC's starting point, that perfect
precision is practically unattainable, then
clause~\ref{en:  AOVProx} of the AOV principle,
relating as it does to  approximate
measurements, must be the one which really matters. 
Clause~\ref{en:  AOVIdeal}, by contrast, relating as
it does to ideal measurements---which is to say
practically unrealizable measurements---must be
regarded as being of negligible importance.  Yet
clause~\ref{en:  AOVIdeal} is the only one on which
their argument  bears.

\begin{figure}[hb]
\begin{picture}(350,100)
\put(45,50){\framebox(60,30){$(\mathbf{e}_1
\cdot \mathbf{L})^2$}}
\put(145,50){\framebox(60,30){$(\mathbf{e}_2
\cdot \mathbf{L})^2$}}
\put(245,50){\framebox(60,30){$(\mathbf{e}_3
\cdot \mathbf{L})^2$}}
\put(0,65){\vector(1,0){23}}
\put(23,65){\line(1,0){22}}
\put(105,65){\vector(1,0){20}}
\put(125,65){\line(1,0){20}}
\put(205,65){\vector(1,0){20}}
\put(225,65){\line(1,0){20}}
\put(305,65){\vector(1,0){23}}
\put(328,65){\line(1,0){22}}
\end{picture}
\caption{
Schematic arrangement for jointly
measuring the observables 
$(\mathbf{e}_r
\cdot
\hat{\mathbf{L}})^2$.  The system
passes through a succession of
analysers, which measure each observable
in turn.}
\label{fig:  SeqMeas}
\end{figure}
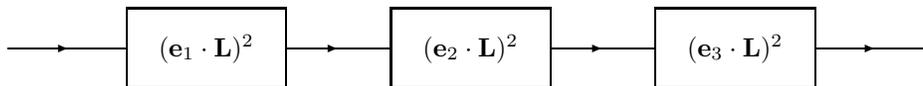

In order to appreciate the force of Mermin's point
it will be helpful to specialise to the standard
example of a spin-1 particle, with angular momentum
$\hat{\mathbf{L}}$.  Kochen and Specker consider joint
measurements of the three commuting projectors 
$(\mathbf{e}_r \cdot \hat{\mathbf{L}})^2$ for
$r=1,2,3$ where
$\mathbf{e}_r$ is any orthonormal triad 
in $\mathbb{R}^3$.  A schematic arrangement for
performing such a measurement using three separate 
analyzers is illustrated in Fig.~\ref{fig:  SeqMeas}.
MKC correctly observe that, in such an arrangement, it
would not practically be possible to align the
analyzers precisely along the three directions
$\mathbf{e}_r$.  In practice one would expect there to
be some uncontrollable errors, so that what are
actually measured are the projections
$(\mathbf{e}'_r \cdot \hat{\mathbf{L}})^2$, where
the triad $\mathbf{e}'_r$ is close, but not exactly
coincident with the triad $\mathbf{e}_r$. 
Nevetheless, MKC assume that the triad 
$\mathbf{e}'_r$ is precisely orthonormal.  Yet it
seems clear that, given that the errors
are random and uncontrollable, and given that the
analyzers are separate instruments, one would 
typically expect there to be some slight departures
from strict orthogonality.  At the least, there are no
evident grounds for assuming the contrary.

MKC assume that the triad
$\mathbf{e}'_r$ must be exactly orthonormal because
they follow  Kochen and Specker in relying on the
principle that it is only sets of commuting
observables which can  jointly be measured with
perfect accuracy.  However, if one relaxes the
requirement that the measurements be perfectly
accurate, then this principle is no longer valid. 
There is  now an extensive literature on the subject of
joint, inexact measurements of non-commuting
observables.  To date, the topics which have received 
most attention are joint  measurements of position and
momentum~\cite{self1a,self1b,PosMom,self1c,self2}, and
joint measurements of the components   of
spin~\cite{spin,self3}.  For recent reviews, and
additional references, the reader may consult Busch
\emph{et al}~\cite{BuschBk}, and
Leonhardt~\cite{LeonBk}.  It should be stressed that
recent advances in the field of quantum optics mean
that such measurements can now be realized, in the
laboratory.  

The purpose of this paper is to investigate the
consequences of extending the analysis of MKC, so as to
include approximate joint measurements of non-commuting
observable.

The paper is in two main parts.  The first part,
comprising 
Sections~\ref{sec:  MeasThy}--\ref{sec:  JMeasTriad},
is concerned with the theory of approximate
measurements.  The discussion in  these sections is
based on ideas previously presented in 
Appleby~\cite{self1a,self1b}
(also see   Appleby~\cite{self1c,self2,self3}).  In our
earlier papers we were concerned with the two special
cases, of approximate joint measurements of position
and momentum~\cite{self1a,self1b,self1c,self2}, and
approximate joint measurements of the components of
spin~\cite{self3}. In  Sections~\ref{sec:  MeasThy}
and~\ref{sec:  POVM} we show how  the same  methods
can be used to analyse  approximate measurements for
any system having a finite dimensional state space
(the extension to the case of a system having an 
infinite dimensional state space is straightforward,
but unnecessary for present purposes).  In
Section~\ref{sec:  Unsharp} we discuss
Uffink's~\cite{Uffink} criticisms of the description
of approximate joint measurement processes which
(unlike the approach taken in this paper) is based on
the concept of an unsharp observable. In
Section~\ref{sec:  JMeasTriad} we specialise the
discussion to the case of approximate joint
measurements of the operators 
$(\mathbf{e}'_r \cdot \hat{\mathbf{L}})^2$ introduced
above, where
$\hat{\mathbf{L}}$ is the angular momentum for a
spin-1 system, and where the triad
$\mathbf{e}'_r$ is not assumed to be orthonormal.

The account in 
Sections~\ref{sec:  MeasThy}--\ref{sec:  JMeasTriad}
is somewhat lengthy.  This is because there
are some subtleties, and serious potential confusions,
which make it necessary to give a  detailed
discussion of the underlying concepts.  
Uffink~\cite{Uffink} (also
see Fleming~\cite{Flem}) has identified some
obscurities in the theory of joint measurements of
non-commuting observables as it is presented by (for
example) Busch
\emph{et al}~\cite{BuschBk}.  It so happens that these
objections have a direct bearing on the questions
addressed in this paper.  One of the  advantages of
our approach  is that Uffink's objections do not apply
to it (the other advantage of our approach being that
it leads to an improved~\cite{self1c} definition of
measurement accuracy).  The importance of this fact
will appear in  Section~\ref{sec:  CKonPOVMs}, where we
consider the argument of  Clifton and 
Kent~\cite{KentB} which  (they claim) ``rule[s] out
falsifications of non-contextual models based on
generalized observables, represented by POV measures''.

In Section~\ref{sec:  context} we apply the
concepts and methods developed in 
Sections~\ref{sec:  MeasThy}--\ref{sec:  JMeasTriad}
to joint measurements of the target observables
$(\mathbf{e}_r
\cdot \hat{\mathbf{L}})^2$, under circumstances where
the triad $\mathbf{e}_r$ is not precisely specified,
so that the observables $(\mathbf{e}'_r \cdot
\hat{\mathbf{L}})^2$ which are actually measured
cannot be assumed to be precisely commuting.  We show
that, if the errors in the alignments of the vectors
$\mathbf{e}_r$ are statistically independent, then it
is possible to prove a modified version of the
Kochen-Specker theorem (a Kochen-Specker theorem for
approximate measurements, as it might be called), from
which it follows that clause~\ref{en:  AOVProx} of the
AOV principle is not satisfied.   This argument 
was originally inspired by some of the points made by
Mermin~\cite{Mermin} (although it appears to us that
our formulation is considerably sharper than Mermin's: 
moreover, Mermin does not remark on the need to assume
that the errors are independent).

The result proved in Section~\ref{sec:  context}
shows that, if the alignment errors are independent,
then the outcome of an approximate measurement must,
in general, be strongly dependent on the particular
manner in which the measurement is carried out.  In
other words, the theory must exhibit a kind of
contextuality.  However, it may be that there are
theories of the MKC type for which the errors are not
independent.  This possibility is discussed in 
Section~\ref{sec:  independence}.  We begin by
remarking that, although it is
conceivable that there exist theories of the type
proposed by MKC for which the errors are not
independent, and which do satisfy all three clauses of
the AOV principle, it would not be 
straightforward actually to prove that this was the
case.  The distribution of errors is
not a feature of the theory which one is free simply
to postulate.  In a complete theory it should be
a consequence of the detailed dynamics of the
interaction between the system, the measuring
apparatus, and the environment.  The models proposed
by MKC are, as they stand, incomplete, since they do
not include any dynamical postulate.  In order to
show that they satisfy clause~\ref{en:  AOVProx} the
AOV principle it would be necessary, first to specify
the dynamical evolution of the hidden variables which
characterise the interacting
system+apparatus+environment composite, and then to
work out the distribution of errors which this
implies.  Moreover, one would need to establish that
the assumption of independence fails in just the way
that is required for clause~\ref{en:  AOVProx} of the
AOV principle  to be satisfied; and one would need to
show that this is the case for every possible system,
and every possible set of measurements.  Such a
program would constitute a highly non-trivial
theoretical undertaking.  We may therefore conclude,
in the first place, that it remains an open question,
whether there exists a hidden variables theory
satisfying clause~\ref{en:  AOVProx} of the AOV
principle.  Whether or not this clause 
\emph{can} be
nullified, it has not been nullified  \emph{yet}.

In the second place it is to be observed that
such a theory,  if it could be constructed,  would
entail the existence of
 a delicately adjusted collaboration
between the ostensibly random fluctuations in the
different parts of a composite apparatus.  In 
Section~\ref{sec:  independence} we argue that this
would itself represent 
a kind of contextuality:  for it would mean that the
fluctuations in each component of
a complex apparatus were, in general, intricately
and inescapably dependent on the overall experimental
context in which that component was employed.  
In other words, a theory of this kind would not
so much eliminate the phenomenon of contextuality,
as shift the locus  of the contextuality, from
the system, onto the fluctuations in the measuring
apparatus.    

Our overall conclusion consequently is that, no matter
how the theoretical postulates are adjusted, 
some kind of contextuality must appear
somewhere.

Finally, in Section~\ref{sec:  CKonPOVMs} we discuss
Clifton and Kent's~\cite{KentB} theorem 2 which is
intended to ``rule out falsifications of
non-contextual models based on generalized
observables, represented by POV measures'' (also
see Kent~\cite{KentA}).
Approximate joint measurements of non-commuting
observables are most conveniently described using a
POVM, and so it may
at first sight seem that Clifton and Kent's theorem 2
contradicts the result proved in Section~\ref{sec: 
context} of this paper.  In fact, this is not the
case, as we show in  Section~\ref{sec:  CKonPOVMs}. 
The reason is connected with the point made in 
Section~\ref{sec:  Unsharp}:  namely, that although
the concept of an approximate measurement involves
the concept of a  POVM, it does not involve the
concept of a new kind of ``generalized observable'',
distinct from the ordinary kind of observable which is
represented by a self-adjoint operator.  We go on to
discuss some other difficulties which arise from the
way in which Clifton and Kent use the concept of a
generalized observable.
\section{Approximate Measurements}
\label{sec:  MeasThy}
The purpose of this section and the one following is
to give a general characterisation of approximate
measurement(s) performed on  a system having a finite
dimensional state space.  The observables being
measured may be commuting or non-commuting.  Our
approach is based on  ideas previously presented in
Appleby~\cite{self1a,self1b}, in connection with
approximate joint measurements of position and momentum
(also see Appleby~\cite{self1c,self2,self3}). 
As
discussed in  Section~\ref{sec:  Unsharp}, our approach
differs from the one taken by many other authors
in that it makes no use of the concept of an unsharp
observable.  This will prove relevant in 
Section~\ref{sec:  CKonPOVMs}, when we discuss
Clifton and Kent's~\cite{KentB} theorem 2.
The basic physical ideas are described in this
section.  The mathematical elaboration in
terms of POVM's is described in Section~\ref{sec: 
POVM}.

An approximate measurement is a measurement which is
less than perfectly accurate.  It follows, that in
order properly to characterise an approximate
measurement  it is necessary  first to arrive at a
satisfactory, quantum mechanical concept of
measurement accuracy.   This is the problem to which
we now turn.  We  begin by considering the accuracy of
an imperfect measurement of a single observable.  We
then extend the discussion to the case of 
simultaneous, imperfect measurements of a set of
several different observables (commuting
or non-commuting).

The ordinary, intuitive
concept of accuracy involves a comparison between
the result of the measurement, and the original
value which the quantity being measured did take,
immediately before the measurement was carried
out.  In a quantum mechanical context this
concept becomes  problematic.  The reason for
this is the very feature of quantum mechanics
which the Kochen-Specker theorem was intended to
establish:  namely, the fact that in quantum
mechanics the concept of ``the original value of
the observable being measured'' is not always
well-defined.  
Of course, one is free to make it 
well-defined, by taking a hidden variables
approach.  However, this way of arriving at a
concept of quantum mechanical accuracy is not
satisfactory because, quite apart from the fact
that it compels one to favour a hidden variables
approach over all the many alternatives, it 
makes the accuracy strongly dependent on which
particular hidden variables theory  is
adopted.  It is arbitrary, in other words.  What
one wants is a concept of accuracy which is (1) a
natural generalization of the classical concept
and  (2) independent of the way in which the
theory is interpreted (so that it is a feature
of quantum mechanics
\emph{as such}, and not simply a feature of this
or that particular interpretation).  In the following
we will present a solution to this problem.

Let us start with the standard,
elementary textbook example of a measurement
process.  Consider a system, with 
finite dimensional state space
$\mathcal{H}_{\mathrm{sy}}$, and
an apparatus, 
with finite dimensional state space
$\mathcal{H}_{\mathrm{ap}}$.  Let   
$\hat{A}$ be a system observable
acting on $\mathcal{H}_{\mathrm{sy}}$, 
and let $\hat{\alpha}$ be a pointer observable
acting on  $\mathcal{H}_{\mathrm{ap}}$.  Suppose
that $\hat{A}$ and $\hat{\alpha}$ have the same
set of eigenvalues $\{a\}$, which for simplicity
we will assume to be non-degenerate.  Let
$\ket{a}_{\mathrm{sy}}$ be the corresponding 
eigenvectors of the operator $\hat{A}$, and 
let $\ket{a}_{\mathrm{ap}}$ be the eigenvectors
of $\hat{\alpha}$.  Let
$\ket{\phi_0}_{\mathrm{ap}}$ be the initial 
``zeroed'' or ``ready'' state of the apparatus,
and let $\sum_{a} c_a \ket{a}_{\mathrm{sy}}$ be
the initial state of the system.  We then obtain
an idealised measurement process by postulating
that the unitary evolution operator
$\hat{U}$ describing the interaction between
system and apparatus is such that
\begin{equation}
  \hat{U}\biggl(\Bigl(
\sum_{a} c_a \ket{a}_{\mathrm{sy}}\Bigr)
\otimes
\ket{\phi_0}_{\mathrm{ap}}\biggr)
=  \sum_{a} c_a \Bigl(\ket{a}_{\mathrm{sy}}
\otimes
\ket{a}_{\mathrm{ap}}\Bigr)
\label{eq:  IdealMeas}
\end{equation}
What makes this a measurement is the fact that it
establishes a correlation between the system and
pointer observables.  What makes it ideal is the fact
that the correlation is, in a certain sense,
perfect.  Specifically:
\begin{enumerate}
\item 
\label{item:  RetIdeal}
The measurement is
retrodictively ideal in the sense that, if the
system was initially
 in the  eigenstate of $\hat{A}$ with
eigenvalue $a$, then there is probability 1
that the recorded value of the pointer observable
will also be
$a$.  Consequently, if the system was prepared in
some unknown eigenstate of $\hat{A}$, the result
of the measurement can be used to retrodict, with
certainty, which particular eigenstate it was.
\item
\label{item:  PreIdeal}
The measurement is predictively ideal in the
sense that, if the pointer observable is recorded
as having the value $a$ immediately after the
measurement, then one can predict, with
probability 1, that a second, immediately
subsequent retrodictively ideal measurement of
$\hat{A}$  will give the same value $a$.
\end{enumerate}
It is easily seen that these two properties, of
retrodictive and predictive ideality, are
independent.  That is, there exist unitary
evolution operators $\hat{U}$ describing
processes which are retrodictively but not
predictively ideal; and operators
$\hat{U}$ describing processes which are
predictively but not retrodictively ideal.

Practically speaking perfection is seldom, if
ever attainable.  Consequently, one does not
expect a real measurement process to be either
retrodictively or predictively ideal.  A more
realistic model of a measurement process is
obtained if, instead of Eq.~(\ref{eq: 
IdealMeas}), we take the evolution to be
described  by
\begin{equation}
  \hat{U}\biggl(\Bigl(
\sum_{a} c_a \ket{a}_{\mathrm{sy}}\Bigr)
\otimes
\ket{\phi_0}_{\mathrm{ap}}\biggr)
=  \sum_{a} c_a \Bigl(\ket{a}_{\mathrm{sy}}
\otimes
\ket{a}_{\mathrm{ap}}\Bigr)
+
\sum_{a,b,d} c_a  \epsilon_{a,bd}
\Bigl(\ket{b}_{\mathrm{sy}}
\otimes
\ket{d}_{\mathrm{ap}}\Bigr)
\label{eq:  ProxMeas}
\end{equation}
where $|\epsilon_{a,bd}|\ll 1$ for all
$a,b,d$.  Of course, this model does not include
all the complications which one might expect to
find in a real measurement process.  A complete
account should allow for the existence of other
apparatus degrees of freedom, additional to
$\hat{\alpha}$.  It should also allow for the
interaction with the environment~\cite{PeresB},
and for the fact that $\hat{A}$ and
$\hat{\alpha}$ may not have exactly the same
spectrum.  However, the model just indicated has
the merit of simplicity, and it will serve to
illustrate the essential ideas.

If the
coefficients 
$\epsilon_{a,bd}$ are sufficiently small, then the
process described by  Eq.~(\ref{eq:  ProxMeas}) may be
regarded as an approximate measurement of
$\hat{A}$:  for, corresponding to the properties
\ref{item:  RetIdeal} and \ref{item:  PreIdeal} above,
we have
\begin{enumerate}
\setcounter{enumi}{2}
\item The measurement is retrodictively good  in the
sense that, if the system was prepared in some
unknown eigenstate of
$\hat{A}$, then the result of the measurement can
be used to retrodict, with a high degree of
confidence, which particular eigenstate it was.
\item 
The measurement is predictively good 
in the sense that, if the pointer observable is
recorded as having the value $a$ immediately after the
measurement, then one can predict, with probability
close to 1, that a second, immediately subsequent
retrodictively ideal measurement of
$\hat{A}$  will give the same value $a$.
\end{enumerate}

We next show how it is possible to quantify the degree
of accuracy of the measurement.  Define
\begin{align*}
  \hat{A}_\mathrm{f} 
& = \hat{U}^{\dagger} \hat{A} \hat{U}
\\
  \hat{\alpha}_\mathrm{f} 
& = \hat{U}^{\dagger} \hat{\alpha} \hat{U}
\end{align*}
$\hat{A}_\mathrm{f}$,
$\hat{\alpha}_{\mathrm{f}}$ are the final
Heisenberg picture observables, defined at the
moment the measurement interaction is completed. 
Let $\hat{A}_{\mathrm{i}}=\hat{A}$ denote the
initial Heisenberg picture system observable,
defined at the moment the measurement interaction
begins.  Define the retrodictive error
operator $\hat{\epsilon}_{\mathrm{i}}$
and predictive error operator 
$\hat{\epsilon}_{\mathrm{f}}$ by
\begin{align}
\hat{\epsilon}_{\mathrm{i}}
& = \hat{\alpha}_{\mathrm{f}}-\hat{A}_{\mathrm{i}}
\label{eq:  Err1Opi}
\\
\hat{\epsilon}_{\mathrm{f}}
& = \hat{\alpha}_{\mathrm{f}}-\hat{A}_{\mathrm{f}}
\label{eq:  Err1Opf}
\end{align}
Let $\mathcal{S}_{\mathrm{sy}}$ denote the unit
sphere $\subset \mathcal{H}_{\mathrm{sy}}$. 
Following the discussion in 
Appleby~\cite{self1b} we now define the maximal
rms error of retrodiction, $\Delta_{\mathrm{ei}}
A$ by
\begin{align}
  \Delta_{\mathrm{ei}} A
& = \biggl( \sup_{\psi \in
           \mathcal{S}_{\mathrm{sy}}}
      \Bigl( \bmat{\psi \otimes \phi_0
                }{\hat{\epsilon}_{\mathrm{i}}^2
                }{\psi \otimes \phi_0
                 }
      \Bigr)
      \biggr)^{\frac{1}{2}}
\label{eq:  Max1RmsRet}
\\
\intertext{and the maximal
rms error of prediction, $\Delta_{\mathrm{ef}}
A$ by}
  \Delta_{\mathrm{ef}} A
& = \biggl( \sup_{\psi \in
           \mathcal{S}_{\mathrm{sy}}}
      \Bigl( \bmat{\psi \otimes \phi_0
                }{\hat{\epsilon}_{\mathrm{f}}^2
                }{\psi \otimes \phi_0
                 }
      \Bigr)
      \biggr)^{\frac{1}{2}}
\label{eq:  Max1RmsPre}
\end{align}

Of these two quantities the predictive error 
$\Delta_{\mathrm{ef}}
A$ is the easier to interpret because 
$\hat{\epsilon}_{\mathrm{f}}$ 
(unlike $\hat{\epsilon}_{\mathrm{i}}$) connects
Heisenberg picture observables defined at the
same instant of time.  Let
$\ket{\psi} \in
\mathcal{H}_{\mathrm{sy}}$ be the (normalised)
initial system state.  Then, reverting to the
Schr\"{o}dinger picture,
\begin{equation*}
  \Bigl(\bmat{\psi \otimes \phi_0
                }{\hat{U}^{\dagger}
                 (\hat{\alpha}-\hat{A})^2
                   \hat{U}
                }{\psi \otimes \phi_0
                 }
  \Bigr)^{\frac{1}{2}}
\le \Delta_{\mathrm{ef}} A
\end{equation*}
from which we see that, the smaller
$\Delta_{\mathrm{ef}} A$, the more closely the
result of a second, immediately subsequent,
retrodictively ideal measurement of 
$\hat{A}$ may be expected to approximate the
result of the (non-ideal) measurement under
discussion.  In particular, if
$\Delta_{\mathrm{ef}} A=0$, then the measurement
is predictively ideal.   It is not
difficult to see that the condition
$\Delta_{\mathrm{ef}} A=0$ is in fact, not only
sufficient, but also necessary for the measurement
to be predictively ideal.  This justifies the
interpretation of
$\Delta_{\mathrm{ef}} A$ as providing a
quantitative indication of the degree of
predictive accuracy.

Let us now consider the interpretation 
of the quantity 
$\Delta_{\mathrm{ei}} A$.
Suppose, to begin with, that the initial system
state $\ket{\psi}$ is an eigenstate of $\hat{A}$
with eigenvalue $a$.  Then
\begin{equation*}
  \Bigl(\bmat{\psi \otimes \phi_0
                }{\hat{U}^{\dagger}
                 (\hat{\alpha}-a)^2
                   \hat{U}
                }{\psi \otimes \phi_0
                 }
  \Bigr)^{\frac{1}{2}}
=
  \Bigl(\bmat{\psi \otimes \phi_0
                }{(\hat{\alpha}_{\mathrm{f}}-a)^2
                }{\psi \otimes \phi_0
                 }
  \Bigr)^{\frac{1}{2}}
\le \Delta_{\mathrm{ei}} A
\end{equation*}
from which we see that, the smaller
$\Delta_{\mathrm{ei}} A$, the more closely
the recorded value of the pointer observable may
be expected to approximate $a$, and the more
accurate the measurement is retrodictively.  In
particular, if
$\Delta_{\mathrm{ei}} A=0$, then the measurement
is retrodictively ideal.  It is not difficult to
see that the condition
$\Delta_{\mathrm{ei}} A=0$ is in fact both
necessary and sufficient for the measurement to
be retrodictively ideal.

It is also possible to say something about the
result of the measurement in the case when the
initial system state $\ket{\psi}$ is not an
eigenstate of $\hat{A}$.  Let $\bar{A}$
and $\Delta A$ denote the initial state
mean and uncertainty:
\begin{align*}
  \bar{A} & = \bmat{\psi}{\hat{A}}{\psi}
\\
  \Delta A 
& = \left( \bmat{\psi}{(\hat{A}-\bar{A})^2}{\psi}
    \right)^{\frac{1}{2}}
\end{align*}
Then the spread of  measured values about the
initial state mean satisfies the inequality
\begin{align}
  \Bigl( \bmat{\psi\otimes\phi_0
            }{(\hat{\alpha}_{\mathrm{f}} -
                 \bar{A})^2
             }{\psi\otimes\phi_0
              }
    \Bigr)^{\frac{1}{2}}
& \le
   \left(\bmat{\psi\otimes\phi_0
            }{(\hat{\alpha}_{\mathrm{f}} -
                 \hat{A}_{\mathrm{i}})^2
             }{\psi\otimes\phi_0
              }
   \right)^{\frac{1}{2}}
\notag
\\
 & \hspace{0.9 in}
   +
   \left(\bmat{\psi\otimes\phi_0
            }{(\hat{A}_{\mathrm{i}} -
                 \bar{A})^2
             }{\psi\otimes\phi_0
              }
   \right)^{\frac{1}{2}}
\notag
\\
 & \le \Delta_{\mathrm{ei}} A + \Delta A
\label{eq:  Ret1NonEig} 
\end{align}
We see from this that there are two components to
the spread of measured values.
$\Delta A$ represents the intrinsic uncertainty
of the initial system state. 
$\Delta_{\mathrm{ei}} A$ represents an upper bound
on the extrinsic uncertainty, attributable to the
noise introduced by the measuring procedure.

These considerations justify the interpretation
of $\Delta_{\mathrm{ei}} A$ as providing a
quantitative indication of the degree of
retrodictive accuracy.

Finally  we note that the
necessary and sufficient condition for the
coefficients
$\epsilon_{a,bd}$ in
Eq.~(\ref{eq:  ProxMeas}) all to be zero (so that
the measurement is completely ideal) is
that
$\Delta_{\mathrm{ei}}A=\Delta_{\mathrm{ef}}A=0$.

Let us now consider a joint
measurement of several different observables.  If
the observables are mutually commuting then the
above discussion generalises in the obvious way. 
However, the point which is important for the
argument of this paper is that it also
generalises, in a manner which is only slightly
less obvious, to the case when the observables
are 
\emph{not} mutually commuting.    It is true that
one cannot make completely ideal joint
measurements of a set of non-commuting
observables.  However, there is nothing to
preclude one from making joint measurements which
are only approximate.

Let $\hat{A}_1, \dots, \hat{A}_n$ be the
non-commuting observables to be measured, acting
on the system state space
$\mathcal{H}_{\mathrm{sy}}$.  Corresponding to
these observables we introduce a set of $n$
pointer observables $\hat{\alpha}_1, \dots,
\hat{\alpha}_n$ acting on the apparatus 
state space $\mathcal{H}_{\mathrm{ap}}$.
We take it that the  observables 
$\hat{\alpha}_1, \dots,
\hat{\alpha}_n$, unlike the observables
$\hat{A}_1, \dots, \hat{A}_n$, are mutually
commuting.  Consequently, their joint eigenvectors
constitute an 
orthonormal basis for
$\mathcal{H}_{\mathrm{ap}}$.  Let 
$\ket{a_1,\dots, a_n}_{\mathrm{ap}}$ denote the
joint eigenvector with eigenvalues $a_1, \dots,
a_n$ (for the sake of simplicity we assume that
the  eigenstates are non-degenerate).  As before,
let $\ket{\phi_0}$ be the initial apparatus
``zeroed'' or ``ready'' state, and let
$\hat{U} \colon \mathcal{H}_{\mathrm{sy}}
\otimes \mathcal{H}_{\mathrm{ap}}
\to \mathcal{H}_{\mathrm{sy}}
\otimes \mathcal{H}_{\mathrm{ap}}$ be the unitary
evolution operator describing the measurement
interaction.

The fact that the observables $\hat{A}_1, \dots,
\hat{A}_n$ are non-commuting means that we cannot
choose a basis for 
$\mathcal{H}_{\mathrm{sy}}$ which consists of
their joint eigenvectors.  However, we can
choose, for each $r$ separately, a basis which
consists of eigenvectors just of 
$\hat{A}_r$.  Let $\ket{a,x}_{r}$ be such a basis
(where $a$ denotes the eigenvalue, and the
additional index
$x$ is to allow for possible degeneracies).  We
may then write
\begin{equation*}
\hat{U}\biggl( \Bigl( \sum_{a,x} c_{ax}
\ket{a,x}_r\Bigr)
\otimes \ket{\phi_0}\biggr)
=
\sum_{b,y,d_1,\dots,d_n} c_{ax} f_{ax;by;d_1\dots
d_n}^{(r)}
\Bigl( \ket{b,y}_r \otimes 
        \ket{d_1,\dots,d_n}_{\mathrm{ap}}
\Bigr)
\end{equation*}
for suitable coefficients 
$f_{ax;by;d_1\dots
d_n}^{(r)}$.  Suppose that, for all $r$, these
coefficients  have the property that
$f_{ax;by;d_1\dots
d_n}^{(r)}$ is small except when
$a=b=d_r$.  Then, comparing this equation with
Eq.~(\ref{eq:  ProxMeas}), we see that, for each
$r$, the pointer $\hat{\alpha}_r$ provides an
approximate measurement of the system
observable $\hat{A}_r$.  This situation may
appropriately be described by saying that
the process provides an approximate joint
measurement of the set of observables
$\hat{A}_1,\dots,\hat{A}_n$.

As in the case of approximate measurements of a
single observable, we may obtain a quantitative
indication of the accuracy by making use of the
Heisenberg picture observables
$\hat{A}_{r\mathrm{f}}=\hat{U}^{\dagger}
\hat{A_r}\hat{U}$,
$\hat{\alpha}_{r\mathrm{f}}=\hat{U}^{\dagger}
\hat{\alpha}_r\hat{U}$,
$\hat{A}_{r\mathrm{i}}=
\hat{A}_r$.  By analogy with 
Eqs.~(\ref{eq:  Err1Opi})
and~(\ref{eq:  Err1Opf}) define
\begin{align*}
  \hat{\epsilon}_{r\mathrm{i}}
& = \hat{\alpha}_{r\mathrm{f}}
     -\hat{A}_{r\mathrm{i}}
\\
  \hat{\epsilon}_{r\mathrm{f}}
& = \hat{\alpha}_{r\mathrm{f}}
     -\hat{A}_{r\mathrm{f}}
\end{align*}
We then obtain $n$ 
maximal rms errors of retrodiction
\begin{align}
  \Delta_{\mathrm{ei}} A_{r}
& = \biggl(\sup_{\psi
\in\mathcal{S}_{\mathrm{sy}}}
\Bigl( \bmat{ \psi\otimes\phi_0
           }{ \hat{\epsilon}_{r\mathrm{i}}^2
           }{ \psi \otimes \phi_0
           }
\Bigr)
  \biggr)^{\frac{1}{2}}
\label{eq:  MaxJRmsRet}
\\
\intertext{and $n$ maximal rms errors of
prediction}
  \Delta_{\mathrm{ef}} A_{r}
& = \biggl(\sup_{\psi
\in\mathcal{S}_{\mathrm{sy}}}
\Bigl( \bmat{ \psi\otimes\phi_0
           }{ \hat{\epsilon}_{r\mathrm{f}}^2
           }{ \psi \otimes \phi_0
           }
\Bigr)
  \biggr)^{\frac{1}{2}}
\label{eq:  MaxJRmsPre}
\end{align}
where $\mathcal{S}_{\mathrm{sy}}$ denotes the unit
sphere in the system state space 
$\mathcal{H}_{\mathrm{sy}}$, as before.

Concerning the interpretation of the
quantities 
$\Delta_{\mathrm{ei}} A_{r}$,
$\Delta_{\mathrm{ef}} A_{r}$ the same analysis
applies to them  as was given for the errors
characterising an approximate measurement of a
single observable, in the paragraphs following 
Eqs.~(\ref{eq:  Max1RmsRet})
and~(\ref{eq:  Max1RmsPre}).  In particular, 
we have, by analogy with
Inequality~(\ref{eq:  Ret1NonEig}),
\begin{equation*}
  \Bigl( \bmat{\psi\otimes\phi_0
            }{(\hat{\alpha}_{\mathrm{rf}} -
                 \bar{A}_r)^2
             }{\psi\otimes\phi_0
              }
    \Bigr)^{\frac{1}{2}}
\le \Delta_{\mathrm{ei}} A_r + \Delta A_r
\end{equation*}
for $r=1,\dots,n$, where 
$\bar{A}_r$ denotes the
initial state mean, and
$\Delta
A_r$ denotes the initial state uncertainty, as in
Inequality~(\ref{eq:  Ret1NonEig}).

Even though the $\hat{A}_r$ are non-commuting, it
may still happen  that there exist states for
which the intrinsic uncertainties
$\Delta A_r$ are all small.  If the retrodictive
errors
$\Delta_{\mathrm{ei}}A_r$ are also small, then
the above inequalities show that there is a high
probability that, for each $r$, the recorded value
of 
$\hat{\alpha}_r$ will be close to the initial
state expectation value $\bar{A}_r$---which
provides a further illustration of the sense in
which the processes under discussion may be
regarded as  approximate joint measurements.
For  examples of  measurement processes to which 
these comments apply, see 
Appleby~\cite{self1a,self1b,self1c,self2,self3}, and
Section~\ref{sec:  JMeasTriad} below.
\section{Approximate Measurements:  POVM}
\label{sec:  POVM}
An approximate measurement is most conveniently
analysed in terms of the corresponding POVM (positive
operator valued measure).  We avoided introducing this
concept at the outset because we wished to establish
that one can give an adequate theoretical description
of approximate measurements whilst remaining wholly
within the framework of the conventional theory, as it
was presented by Dirac~\cite{Dirac} and von
Neumann~\cite{VNeu}.  In particular, we wished to
establish that one can introduce the concept of an
approximate measurement, without being thereby 
compelled  to introduce any unconventional, unsharp or
generalized observables. However, it is certainly true
the concept of a POVM represents a powerful
mathematical tool.  Consequently, having
established that it is not anything more
than a tool (at least in the present context), it is
appropriate to indicate how the maximal rms errors
defined in Section~\ref{sec:  MeasThy} can be
expressed in terms of this construct.

As in the last section, we consider a measurement
of $n$ non-commuting observables $\hat{A}_1,
\dots,
\hat{A}_n$ acting on the system state space
$\mathcal{H}_{\mathrm{sy}}$.  The system is
coupled to $n$ commuting pointer observables
$\hat{\alpha}_1, \dots, \hat{\alpha}_n$ acting on
the apparatus state space
$\mathcal{H}_{\mathrm{ap}}$.  Let 
$\ket{a_1, \dots, a_n}$ be the
joint eigenvector of $\hat{\alpha}_1, \dots,
\hat{\alpha}_n$ with eigenvalues $a_1, \dots,
a_n$ (which, for simplicity, we assume to be
non-degenerate).  Let
$\ket{\phi_0}$ be the initial apparatus state,
and let
$\hat{U}$ be the unitary evolution operator
describing the measurement interaction. Let
$\ket{m}$ be any orthonormal basis for the
system space $\mathcal{H}_{\mathrm{sy}}$. Define,
for each
$n$-tuplet
$a_1,\dots,a_n$,
\begin{equation}
\hat{T}_{a_1, \dots, a_n}
=\sum_{m,m'} 
   \bigl(\bra{m} \otimes
\bra{a_1,\dots,a_n}
   \bigr) \hat{U}
   \bigl( \ket{m'} \otimes 
          \ket{\phi_0}
   \bigr) \ket{m}\bra{m'}
\label{eq:  TDef}
\end{equation}
Unlike $\hat{U}$, which acts on the product 
space $\mathcal{H}_{\mathrm{sy}}\otimes
\mathcal{H}_{\mathrm{ap}}$, the operators 
$\hat{T}_{a_1, \dots, a_n}$ act just on the
system space $\mathcal{H}_{\mathrm{sy}}$.

Let
\begin{equation}
  \hat{E}_{a_1,\dots,a_n}
= \hat{T}^{\dagger}_{a_1,\dots,a_n}
   \hat{T}_{a_1,\dots,a_n}^{\vphantom{\dagger}}
\label{eq:  POVMGenCase}
\end{equation}
It is easily verified that 
$\hat{E}_{a_1,\dots,a_n}$ is the POVM describing
the measurement outcome.  In other words, the
probability that, immediately after the
measurement, the $n$ pointers will be recorded
as having the values $a_1, \dots, a_n$ is
\begin{equation*}
  p_{a_1,\dots,a_n}
= \bmat{\psi}{\hat{E}_{a_1,\dots,a_n}}{\psi}
\end{equation*}
where $\ket{\psi}$ is the initial state of the
system, immediately before the measurement.

It is also convenient to define
\begin{equation*}
  \hat{E}_{a_r}^{(r)}
= \sum_{a_1,\dots,a_{r-1},a_{r+1},\dots,a_{n}}
  \hat{E}_{a_1,\dots,a_n}
\end{equation*}
where the summation is over every index 
except for $a_r$.
$\hat{E}_{a_r}^{(r)}$ is the POVM describing the
outcome of the measurement just of
$\hat{A}_{r}$, which is obtained by ignoring the
other $n-1$ pointer readings.  Thus, the
probability that the $r^{\mathrm{th}}$ pointer
reading will be $a_r$ is given by
\begin{equation*}
  p_{a_r}^{(r)} =
\bmat{\psi}{\hat{E}_{a_r}^{(r)}}{\psi}
\end{equation*}
Starting from the definition of  
Eq.~(\ref{eq:  MaxJRmsRet}) it is not difficult
to show that the $r^{\mathrm{th}}$ retrodictive
error
$\Delta_{\mathrm{ei}}A_r$ is given by
\begin{equation*}
  \Delta_{\mathrm{ei}}A_r
= \biggl( \sup_{\psi\in \mathcal{S}_{\mathrm{sy}}}
      \Bigl(\bmat{ \psi
                }{\sum_{a_r}
                  (\hat{A}_r-a_r)
                  \hat{E}^{(r)}_{a_r}
                  (\hat{A}_r-a_r)
                }{\psi
                }
      \Bigr)
   \biggr)^{\frac{1}{2}}
\end{equation*}
or, equivalently,
\begin{equation}
  \Delta_{\mathrm{ei}}A_r
= \biggl(\Bigl\|\sum_{a_r}
                  (\hat{A}_r-a_r)
                  \hat{E}^{(r)}_{a_r}
                  (\hat{A}_r-a_r)
          \Bigr\|\biggr)^{\frac{1}{2}}
\label{eq:  RetErrGenJ}
\end{equation}
where $\| \cdot \|$ denotes the operator norm. 
Similarly, the $r^{\mathrm{th}}$ predictive error
may be expressed
\begin{equation}
  \Delta_{\mathrm{ef}}A_r
= \biggl(\Bigl\|\sum_{a_1,\dots,a_n}
                 \hat{T}^{\dagger}_{
																		a_1,\dots,a_n}
                  (\hat{A}_r-a_r)^2
       \hat{T}^{\vphantom{\dagger}}_{
																		a_1,\dots,a_n}
          \Bigr\|\biggr)^{\frac{1}{2}}
\label{eq:  PreErrGenJ}
\end{equation}
\section{Approximate Measurements and ``Unsharp
Observables''}
\label{sec:  Unsharp}
As we stressed earlier, the approach described in
Sections~\ref{sec:  MeasThy} 
and~\ref{sec:  POVM} differs from the approach of many
other authors in that we make no use of the concept of
an ``unsharp observable'', or of what Clifton and
Kent~\cite{KentB} refer to as a ``generalized
observable''.  In this section we discuss
Uffink's~\cite{Uffink} criticisms (also see
Fleming~\cite{Flem}) of this way of describing joint
measurements of non-commuting observables.  The
discussion will prove relevant in Section~\ref{sec: 
CKonPOVMs}, where we consider Clifton and
Kent's~\cite{KentB} argument  ``to rule out
falsifications of non-contextual models based on
generalized observables, represented by POV measures''.

Historically, work on the application of
POVM's to the theory of measurement has been strongly
influenced by the fact that, from a
mathematical point of view, the concept of a POVM
(positive operator valued measure) is a
generalization of the concept of a PVM (projection
valued measure).   There consequently arose   the idea
that, since observables of the ordinary, orthodox kind
are represented by PVMs, therefore a POVM which is
not also a PVM must represent an observable of a
different, unorthodox kind. 

If one takes such  a view,    then
one has to suppose that what  would
naturally be regarded as an approximate measurement of
(for example) position, is in fact a (non-approximate?)
measurement of something else---unorthodox, or
generalized, or unsharp position as it might be called.
This way of thinking is certainly at variance with
our ordinary intuitions.  It would, for instance, not
normally be argued that a ruler cannot be used to
measure length properly so-called, but only generalized
length.  However, this objection is perhaps not
crucial, for one does not  expect quantum mechanical
concepts necessarily to accord with classical
intuition.  Nevertheless, there are some pertinent
questions regarding the interpretation of generalized
observables which need to be answered if the concept
is to be acceptable.  As Uffink puts it:
``one would naturally like to know \emph{what} is
being measured in a measurement of an unorthodox
observable'' (his emphasis).

The difficulty  becomes particularly acute when it is
 approximate joint measurements of non-commuting
observables which are in question.  Proponents of the
concept of an unsharp observable argue that,
although (for example) the orthodox  position
and momentum observables cannot jointly be measured,
there exists a different pair of 
unorthodox, ``unsharp'' observables
which are jointly measurable.  As Uffink points out,
the problem with this approach is that, rather than
solving the original problem  (the problem
of making a joint measurement of a pair of
\emph{orthodox} observables), it merely presents us
with a solution to a new, ostensibly
quite different problem (the problem
of making a joint measurement of a pair of
\emph{unorthodox} observables).  Advocates of the
approach attempt to deal with this problem by arguing
that the unorthodox observables 
which one actually measures are related to (are
unsharp versions of) the orthodox observables 
which one would like to measure in such a way that, by
making a (non-approximate?) measurement of the
former, one acquires approximate information
regarding the latter.  However, Uffink has identified
some  problems with this
idea~\cite{Uffink,Flem}.

It appears to us that the source of the difficulty
lies in the concept of an unsharp observable which,
at least so far as approximate measurements are
concerned,
 adds a wholly unnecessary level of
complication  to the  problem.  In classical
physics there is no need to introduce the
concept of an ``unsharp''  quantity, and then 
attempt to show that, by measuring that, one
gains approximate information about the ordinary
quantity in which one is really interested.  It turns
out that there is no need to introduce such
intermediate quantities in quantum physics
either---as we showed in   Sections~\ref{sec: 
MeasThy} and~\ref{sec:  POVM} where (using
ideas previously presented in  
Appleby~\cite{self1a,self1b,self1c,self2,self3}) we
described approximate quantum mechanical
measurements  directly,  without any unnecessary
detours, as measurements of the self-same (orthodox)
observables concerning which approximate information
is sought.

As was shown    in  
Section~\ref{sec:  POVM}, the concept of a POVM plays
an important role in our analysis.  However, its role
is simply that of a powerful mathematical construct,
which can be used to describe the outcome of an
approximate measurement.  It is not taken to
represent a new kind of observable, distinct
from the observable  one is trying
approximately  to measure.

It should be stressed that the above discussion
only applies to approximate measurements of
(orthodox) observables.  In other contexts we would
agree that the orthodox identification of
``observable'' with ``self-adjoint operator'' is 
too restrictive---as appears from the fact that, if
this identification is correct, then phase and time are
not observables (see, for example, Busch \emph{et
al}~\cite{BuschB}, Pegg and Barnett~\cite{Pegg},
Bu\v{z}ek \emph{et al}~\cite{Buzek}, Oppenheim
\emph{et al}~\cite{Opp}, Egusquiza and
Muga~\cite{Muga}, and references cited therein).  It
is also clearly true that the concept of a POVM plays
an important role in the problem of arriving at a
suitably extended concept of a physical
observable.  We only wish to point out that the
question is not straightforward, and that a
simple identification of the concept of a POVM with the
concept of a generalized observable may be productive
of confusion.  Some of the  pitfalls appear from
the discussion in Uffink's paper.  Others will appear
from the discussion in Section~\ref{sec:  CKonPOVMs}.
\section{Approximate Joint Measurements
of the Projections $(\mathbf{e}_r 
\cdot \hat{\mathbf{L}})^2$}
\label{sec:  JMeasTriad}
We now specialise the
theory presented in Sections~\ref{sec:  MeasThy}
and~\ref{sec:  POVM} to the
case which will be discussed in the next two sections,
of an approximate joint measurement of the projections
$(\mathbf{e}_1 
\cdot \hat{\mathbf{L}})^2$,
$(\mathbf{e}_2 
\cdot \hat{\mathbf{L}})^2$,
$(\mathbf{e}_3 
\cdot \hat{\mathbf{L}})^2$ where
$\hat{\mathbf{L}}$ is the angular momentum
operator for a spin 1 system,
and where the unit vectors $\mathbf{e}_1$,
$\mathbf{e}_2$, $\mathbf{e}_3$ are approximately,
but perhaps not exactly orthogonal.

Let us start by considering an exact measurement
of the single projection $\hat{P} =(\mathbf{n}
\cdot
\hat{\mathbf{L}})^2$, for an arbitrary unit
vector $\mathbf{n}$.  The system state space
$\mathcal{H}_{\mathrm{sy}}$ is thus
3-dimensional.  To measure $\hat{P}$ we couple
the system to a single pointer observable
$\hat{\alpha}$ which has the two (non-degenerate)
eigenvalues $0$ and $1$. The apparatus  state
space 
$\mathcal{H}_{\mathrm{ap}}$ is thus
2-dimensional.  Let $\ket{0}$,
$\ket{1}$ be the eigenvectors of
$\hat{\alpha}$ with eigenvalues $0$ and $1$
respectively.  Let $\hat{\sigma}\colon
\mathcal{H}_{\mathrm{ap}}
\to
\mathcal{H}_{\mathrm{ap}}$ be the
operator defined by
\begin{equation}
  \hat{\sigma} \ket{0}
 = - i \ket{1}
 \hspace{1.0 in}
  \hat{\sigma} \ket{1}
 = i \ket{0}
\label{eq:  SigDef}
\end{equation}
Let $\hat{U}\colon
\mathcal{H}_{\mathrm{sy}}\otimes
\mathcal{H}_{\mathrm{ap}}
\to
\mathcal{H}_{\mathrm{sy}}\otimes
\mathcal{H}_{\mathrm{ap}}$ be the unitary
 operator defined by
 \begin{equation*}
\hat{U}
= \exp\left[i \frac{\pi}{2} 
            \hat{P}  \hat{\sigma}
      \right]
= (1-\hat{P}) + i
\hat{P}\hat{\sigma}
\end{equation*}
Let the initial apparatus state
 be $\ket{\phi_0}
=\ket{0}$.  Then
\begin{equation*}
  \hat{U} \bigl(\ket{\psi}\otimes
                 \ket{0}\bigr)
=
  \begin{cases} \ket{\psi}\otimes
          \ket{0}
          \hspace{0.5 in}
          \text{if $\hat{P}\ket{\psi}=0$}
          \\
          \ket{\psi}\otimes
          \ket{1}
          \hspace{0.5 in}
          \text{if $\hat{P}\ket{\psi}=
            \ket{\psi}$}
  \end{cases}
\end{equation*}
from which we see that 
$\hat{U}$ describes a completely ideal measurement
of $\hat{P}$.

In order to obtain a joint measurement of the
three projections
$\hat{P}_1=(\mathbf{e}_1 
\cdot \hat{\mathbf{L}})^2$,
$\hat{P}_2=(\mathbf{e}_2 
\cdot \hat{\mathbf{L}})^2$,
$\hat{P}_3=(\mathbf{e}_3 
\cdot \hat{\mathbf{L}})^2$ we can chain
together three 
ideal measurements of the kind just described, so
that $\hat{P}_1$ is measured first, $\hat{P}_2$
second and $\hat{P}_3$ third, as illustrated in
Figure~\ref{fig:  SeqMeas}.  We then have three
commuting pointer observables $\hat{\alpha}_1$,
$\hat{\alpha}_2$, $\hat{\alpha}_3$ acting on the
6-dimensional space $\mathcal{H}_{\mathrm{ap}}$.
Let $\ket{\alpha_1,\alpha_2,\alpha_3}$ be
the joint eigenvector of $\hat{\alpha}_1$,
$\hat{\alpha}_2$, $\hat{\alpha}_3$ with eigenvalues
$\alpha_1, \alpha_2, \alpha_3$.
The unitary operator
describing the measurement interaction is
\begin{equation}
\hat{U}=
\hat{U}_3 \hat{U}_2 \hat{U}_1
= 
\bigl((1-\hat{P}_3) + i \hat{P}_3 \hat{\sigma}_{3}
\bigr)
\bigl((1-\hat{P}_2) + i \hat{P}_2 \hat{\sigma}_{2}
\bigr)
\bigl((1-\hat{P}_1) + i \hat{P}_1 \hat{\sigma}_{1}
\bigr)
\label{eq:  UJTriad}
\end{equation}
where the operators $\hat{\sigma}_{r}$ are
defined by
\begin{align*}
\hat{\sigma}_1 \ket{\alpha_1,\alpha_2,\alpha_3}
& = (-1)^{\bar{\alpha}_1} i
\ket{\bar{\alpha}_1,\alpha_2,\alpha_3}
\\
\hat{\sigma}_2 \ket{\alpha_1,\alpha_2,\alpha_3}
& = (-1)^{\bar{\alpha}_2} i
\ket{\alpha_1,\bar{\alpha}_2,\alpha_3}
\\
\hat{\sigma}_3 \ket{\alpha_1,\alpha_2,\alpha_3}
& = (-1)^{\bar{\alpha}_3} i
\ket{\alpha_1,\alpha_2,\bar{\alpha}_3}
\end{align*}
and where we have employed the notation
$\bar{0}=1, \bar{1}=0$.
Referring to Eq.~(\ref{eq:  TDef}) we see that
\begin{align*}
\hat{T}_{\alpha_1 \alpha_2 \alpha_3}
=\sum_{m,m'} 
   \bigl(\bra{m} \otimes
\bra{\alpha_1, \alpha_2, \alpha_3}
   \bigr) \hat{U}
   \bigl( \ket{m'} \otimes 
          \ket{0,0,0}
   \bigr) \ket{m}\bra{m'}
\end{align*}
where $\ket{m}$ is any orthonormal basis
for $\mathcal{H}_{\mathrm{sy}}$.  Defining
$\hat{P}^{(0)}_{r}=1-\hat{P}_r$,
$\hat{P}^{(1)}_{r}=\hat{P}_r$ this becomes
\begin{equation*}
  \hat{T}_{\alpha_1 \alpha_2 \alpha_3}
= \hat{P}_{3}^{(\alpha_3)} \hat{P}_{2}^{(\alpha_2)}
  \hat{P}_{1}^{(\alpha_1)}
\end{equation*}
Using Eq.~(\ref{eq:  POVMGenCase}), and the fact
that the $\hat{P}^{(\alpha_3)}_{3}$ are projections,
the POVM describing the measurement outcome is
\begin{equation*}
\hat{E}_{\alpha_1 \alpha_2 \alpha_3}
=
  \hat{P}_{1}^{(\alpha_1)} \hat{P}_{2}^{(\alpha_2)}
\hat{P}_{3}^{(\alpha_3)} \hat{P}_{2}^{(\alpha_2)}
  \hat{P}_{1}^{(\alpha_1)}
\end{equation*}
We may assume that the  basis in
$\mathbb{R}^3$ has been chosen in such a way that
\begin{equation*}
\mathbf{e}_1 
= \begin{pmatrix}
			  1 \\ 0 \\ 0	
  \end{pmatrix}
\hspace{0.5 in}
\mathbf{e}_2 
= \begin{pmatrix}
			  \sin \psi \\ \cos \psi \\ 0	
  \end{pmatrix}
\hspace{0.5 in}
\mathbf{e}_3 
= \begin{pmatrix}
			  \sin \theta \cos \phi 
     \\ \sin \theta \sin \phi 
     \\ \cos \theta	
  \end{pmatrix}
\end{equation*}
where the angles $\psi, \theta$ 
(but not necessarily $\phi$) are  small.  There
is no loss of generality in assuming that  the
basis is right handed (since we are free to
adjust the signs of the $\mathbf{e}_r$). 
Making this assumption, and working to second
order in $\theta,\psi$, we find (after some
rather lengthy algebra)
\begin{align*}
\hat{E}_{111}
& \approx
\psi^2\,  (1-\hat{L}_2^2)+\theta^2 \,
(1-\hat{L}_3^2)
  - \theta \psi \cos \phi\, 
 \{\hat{L}_2,\hat{L}_3\}
\\
\hat{E}_{110}
& \approx
(1-\theta^2  )\, 
(1-\hat{L}_3^2)+\theta \psi \cos \phi\, 
 \{\hat{L}_2,\hat{L}_3\}
\\
\hat{E}_{101}
& \approx
(1-\theta^2 \sin^2 \phi-\psi^2 )\, 
(1-\hat{L}_2^2)
\\
\hat{E}_{011}
& \approx
(1-\theta^2 \cos^2 \phi-\psi^2 )\, 
(1-\hat{L}_1^2)
\\
\hat{E}_{100}
& \approx
\theta^2 \sin^2 \phi\,  (1-\hat{L}_2^2)
\\
\hat{E}_{010}
& \approx
\theta^2 \cos^2 \phi\,  (1-\hat{L}_1^2)
\\
\hat{E}_{001}
& \approx
\psi^2\,  (1-\hat{L}_1^2)
\\
\hat{E}_{000}
& \approx
0
\end{align*}
where $\{ \hat{L}_2, \hat{L}_3\}$ denotes the 
anti-commutator.  

$\bigl< \hat{E}_{\alpha_1 \alpha_2 \alpha_3} 
\bigr>$ is the
probability that the measurements of $\hat{P}_1$,
$\hat{P}_2$, $\hat{P}_3$
will give the values
$\alpha_1$,  $\alpha_2$,  $\alpha_3$ respectively.  
If
$\psi=\theta=0$ then
$\hat{E}_{\alpha_1 \alpha_2 \alpha_3}$ 
reduces to the 
PVM (projection valued measure)
\begin{equation*}
  \hat{E}_{110}
 =
1-\hat{P}_3
\hspace{0.5 in}
\hat{E}_{101}
 = 
1-\hat{P}_2
\hspace{0.5 in}
\hat{E}_{011}
 =
1-\hat{P}_1
\end{equation*}
\begin{equation*}
 \hat{E}_{111}=\hat{E}_{100}
 =\hat{E}_{010}=\hat{E}_{001} 
 =\hat{E}_{001}=0
\end{equation*}
and the probability of the outcome of the
measurement  being one of the ``illegal''
combinations
$111$, $100$, $010$, $001$, $000$
is zero.  If, however, $\theta$, $\psi$ are not
both $=0$, then the probability of obtaining one
of these combinations, though small, is not
exactly zero---as was to be expected.

Using Eq.~(\ref{eq:  RetErrGenJ}) we obtain
(after some algebra) the following
expressions for the retrodictive errors, to
lowest order in $\psi$, $\theta$:
\begin{align}
  \Delta_{\mathrm{ei}} P_1 & \approx 0 
\label{eq:  RetErrP1}  
\\
  \Delta_{\mathrm{ei}} P_2 & \approx |\psi| \\
  \Delta_{\mathrm{ei}} P_3 & \approx |\theta|
\intertext{For the sake of completeness we also
give the formulae for the predictive errors, to
lowest order in $\psi$, $\theta$:}
  \Delta_{\mathrm{ef}} P_1 
& \approx \bigl(2(\psi^2+\theta^2 \cos^2 
     \phi)\bigr)^{\frac{1}{2}} \\
  \Delta_{\mathrm{ef}} P_2 
& \approx |\sqrt{2}\theta \sin \phi| \\
  \Delta_{\mathrm{ef}} P_3 & \approx 0 
\label{eq:  PreErrP3} 
\end{align}
It is not difficult to see that the equalities
$\Delta_{\mathrm{ei}} P_1=\Delta_{\mathrm{ef}}
P_3=0$ are actually exact.  This is because the
joint measurement is constructed by stringing
together a sequence of measurements which are
individually ideal.  The errors are entirely
attributable to the disturbance of the system
caused by the successive measurements.  The
measurement of
$\hat{P}_1$ comes first, there has been no
preceding measurement to alter the state of the
system, and so it is retrodictively ideal:  
which is why
$\Delta_{\mathrm{ei}} P_1=0$. 
The measurement of $\hat{P}_3$ comes
last, there is no subsequent measurement to alter
the state of the system, and so it
is predictively ideal:  which is why
$\Delta_{\mathrm{ef}} P_3=0$.

We have assumed that the measurements of the three
operators $\hat{P}_r$ are performed sequentially, one
after the other, because that is the easiest case to
analyse.  However, it is perfectly possible to apply
these methods to cases where the three measurements
are performed all at once (so to speak).  For
instance, one might consider the evolution described
by the Hamiltonian
\begin{equation*}
  \hat{H} = \hat{H}_{\mathrm{sy}}
+ \hat{H}_{\mathrm{ap}} + \hat{H}_{\mathrm{meas}}
\end{equation*}
$\hat{H}_{\mathrm{sy}}$ and
$\hat{H}_{\mathrm{ap}}$ are the  Hamiltonians
describing the free evolution of the system and
apparatus respectively.  $\hat{H}_{\mathrm{meas}}$ is
the time-dependent  Hamiltonian describing the
measurement interaction, given by
\begin{equation*}
\hat{H}_{\mathrm{meas}}
=-\frac{h}{4} f(t) 
  \left( \hat{P}_1 \hat{\sigma}_1 +
              \hat{P}_2 \hat{\sigma}_2 +
             \hat{P}_3 \hat{\sigma}_3 
  \right)
\end{equation*}
where $f$ is a ``bump'' function, which is zero
outside the short time interval $[0,\tau]$, and which 
satisfies the normalisation condition
$\int_0^\tau dt f(t) =1$.  If $\tau$ is
sufficiently small then the unitary operator
describing the evolution between $t=0$ and
$t=\tau$ is approximately given by
\begin{equation*}
\hat{U}'
\approx
\exp\left[ i \frac{\pi}{2}
  \left( \hat{P}_1 \hat{\sigma}_1 +
              \hat{P}_2 \hat{\sigma}_2 +
             \hat{P}_3 \hat{\sigma}_3 
  \right)
\right]
\end{equation*}
If the vectors $\mathbf{e}_r$ are exactly
orthonormal, then $\hat{U}'$ coincides with 
the operator $\hat{U}$ given by
Eq.~(\ref{eq:  UJTriad}).  Otherwise it does not.
However, it is easily seen that it still describes an
approximate joint measurement of the operators
$\hat{P}_r$.

Every measurement occupies a finite time interval. 
There is no difference in principle between
joint measurements which are performed sequentially,
so that the measurement of each individual observable
is allotted its own individual time-slice; and joint
measurements which are performed contemporaneously,
so that the measurement of each individual observable
takes up the whole of the time which is allotted to
all.
\section{A Modified Kochen-Specker Argument}
\label{sec:  context}
We now apply the concepts and methods developed in the
last four sections to the questions posed in the
Introduction.
We consider a spin 1 system, with angular momentum
$\hat{\mathbf{L}}$; and we consider the problem of
making a joint measurement of the observables 
$(\mathbf{e}_r \cdot \hat{\mathbf{L}})^2$, for some
 triad $\mathbf{e}_r$.

The argument in this section
was originally inspired by some of the points made by
Mermin~\cite{Mermin}.  However, it appears to us that
our formulation is considerably sharper than Mermin's. 
Moreover, Mermin does not remark on the need to assume
that the alignment errors are statistically independent
(see below).
 
Before proceeding further, it
will be helpful  to
introduce some terminology.  Suppose that
an analyzer is
designed to measure the  observable 
$(\mathbf{n} \cdot \hat{\mathbf{L}})^2$ but, due to
the imprecision in the alignment of the analyzer,
does in fact measure the slightly different
observable $(\mathbf{n}' \cdot \hat{\mathbf{L}})^2$,
where $\mathbf{n}$ and $\mathbf{n}'$ are both unit
vectors. Then we will refer to
$(\mathbf{n}
\cdot
\hat{\mathbf{L}})^2$ as the target observable, and to
the unit vector
$\mathbf{n}$ as the target alignment; while
$(\mathbf{n}' \cdot
\hat{\mathbf{L}})^2$ will be referred to as the 
actual observable, and $\mathbf{n}'$ as the actual
alignment.

Kochen and Specker make no allowance for the
imprecision in any real measurement procedure.  They
consequently assume that the measurements they
consider are all ideal (in the sense explained in 
Section~\ref{sec:  MeasThy}), and
they assume that the actual observables  exactly
coincide with the target observables.  They further
assume that, in a measurement of the three observables
$(\mathbf{e}_r \cdot \hat{\mathbf{L}})^2$, the
vectors
$\mathbf{e}_r$ can be any orthonormal triad contained
in the real unit 2-sphere, $S_2$.

MKC, by contrast, recognise that the imprecision of
any real experimental procedure means that the actual
alignments
$\mathbf{e}'_r$ may be slightly different from the
target alignments
$\mathbf{e}_r$.  This permits them to make the crucial
postulate, that the actual alignments are constrained
to lie in a proper, dense subset $S'_2 \subset S_2$. 
However, MKC continue to assume
that the triad $\mathbf{e}'_r$ is precisely
orthonormal, and that the measurements of the
observables $(\mathbf{e}'_r \cdot \hat{\mathbf{L}})^2$
are all ideal.  As we discussed in the Introduction,
these assumptions are unduly restrictive.   The
argument of MKC has little force unless it can be
extended to the case when the 
triad $\mathbf{e}'_r$ is not precisely
orthonormal and when, in consequence, the measurements
of the (non-commuting) observables $(\mathbf{e}'_r
\cdot
\hat{\mathbf{L}})^2$ are not ideal.  This is the
problem we now address.

The argument which follows does not depend on any
assumption regarding the specific manner in which  the
measurements of the  observables $(\mathbf{e}'_r \cdot
\hat{\mathbf{L}})^2$ are performed. 
The measurements could be performed sequentially, by
three separate analyzers, as illustrated in 
Fig.~\ref{fig:  SeqMeas}.  However, the argument
applies equally well to the case when the measurements
are performed contemporaneously, by a single piece
of apparatus, as discussed at the end of 
Section~\ref{sec:  JMeasTriad}.

In order to proceed it is necessary to make a
definite hypothesis as to the distribution of actual
alignments corresponding to a given target
alignment.  The most straightforward hypothesis, and
the assumption on which the argument of this section
will be based, is that the actual alignments are
distributed randomly.  We will further assume
that, in the case of an apparatus  which is designed to
measure several different target observables,
the  actual observables are distributed
independently.  In other words, we assume that,
for each possible target alignment $\mathbf{n}$,
there is a probability measure $\mu_{\mathbf{n}}$
defined on the set $S'_2$ (\emph{i.e.}, the set to
which MKC postulate that the actual alignment must
belong) such that, in a measurement of the target
observables 
$(\mathbf{e}_1 \cdot \hat{\mathbf{L}})^2$,
$(\mathbf{e}_2 \cdot \hat{\mathbf{L}})^2$,
$(\mathbf{e}_3 \cdot \hat{\mathbf{L}})^2$,
 the probability that the actual alignments
$\mathbf{e}'_1$,
$\mathbf{e}'_2$,
$\mathbf{e}'_3$
lie in the set $A_1 \times A_2 \times A_3
\subseteq S'_2 \times S'_2 \times S'_2$
is $\mu_{\mathbf{e}_1}(A_1)
\mu_{\mathbf{e}_2}(A_2)\mu_{\mathbf{e}_3}(A_3)$.

The probability measure $\mu_{\mathbf{n}}$
depends, not only on the vector $\mathbf{n}$, but
also on the construction of the apparatus.  An
apparatus which was constructed differently, so
as to permit the alignments to be fixed more
precisely, would have a different associated
probability distribution.

The argument of this section is crucially
dependent on the assumption that the alignment errors
are statistically independent.  We will consider the
hypothesis that the distributions are not independent
in Section~\ref{sec:  independence}.

We will denote the set of possible target
alignments $S''_2$.  
The question arises:  what, exactly, is this 
set?  What conditions must 
a vector satisfy in order to be a possible 
target alignment?  One could argue that
\emph{every} vector $\in S_2$ is a possible
target alignment.  However, it might be thought that
this view would be too extreme.  The target
observable is the observable which the analyzer is
designed to measure; and it may
reasonably be argued that it is possible to design
an instrument to measure some observable if and only
if it is  possible unambiguously to describe that
observable.  We will accordingly take the
view that
$S''_2$, the set of possible target alignments, 
 consists of   those unit vectors $\in
S_2$ which are finitely specifiable---that is,
which can be specified in standard mathematical
notation,  by means of a string consisting of 
finitely many characters. 

The set $S''_2$ so
defined is countable, like the set
$S'_2$.  However, unlike the set $S'_2$, 
the set $S''_2$ includes Kochen-Specker (KS)
uncolourable sets.  For instance, it includes the
uncolourable set given by Kochen and
Specker~\cite{Koch} themselves, and the one
given by Peres~\cite{Peres,PeresC} (the vectors
belonging to these sets manifestly are  finitely
specifiable, since the authors explicitly do so
specify them).  It follows that $S''_2$ is itself
KS uncolourable.

The fact that $S''_2$ is KS uncolourable is crucial.
It opens the way to a modified version
of the KS theorem.

Now consider a valuation $f\colon
S'_2\to\{0,1\}$.  The fact that $S'_2$ is
countable means that $f$ is automatically 
$\mu_{\mathbf{n}}$-measurable, for every
$\mathbf{n}\in S''_2$.  Consequently, we may
define for each $\mathbf{n}\in S''_2$,
\begin{equation*}
p(\mathbf{n})
= \mu_{\mathbf{n}} \bigl(\{\mathbf{n}'\in S'_2: 
f(\mathbf{n}')=1\}\bigr)
\end{equation*}
$p(\mathbf{n})$ is the probability that, if the
analyzer is designed to measure the target 
observable $\hat{P}_{\mathbf{n}}$, then the
vector $\mathbf{n}'\in S'_2$ characterising the
actual alignment of the analyzer will have
$f$-value $=1$.

Using the function $p(\mathbf{n})$ we next
define an induced valuation
$\tilde{f}\colon S''_2
\to \{0,1\}$ by
\begin{equation*}
\tilde{f}(\mathbf{n}) =
\begin{cases}
  0 \hspace{0.5 in} \text{if $p(\mathbf{n})<0.5$}
\\ 1 \hspace{0.5 in} \text{if
$p(\mathbf{n})\ge0.5$}
\end{cases}
\end{equation*}
The  valuation $f$ is defined on the set 
$S'_2$, which is KS colourable.  However, the
induced valuation
$\tilde{f}$ is defined on $S''_2$
which, as we have seen, is KS uncolourable.  It
follows that there must exist an orthonormal
triad $\mathbf{e}_1,\mathbf{e}_2,\mathbf{e}_3\in
S''_2$ which
$\tilde{f}$-evaluates to
 one of the ``illegal'' combinations $111$,
$100$, $010$, $001$, $000$.  

Let $\mathbf{e}_1,\mathbf{e}_2,\mathbf{e}_3\in
S''_2$ be such a triad, and suppose that 
the sequence of three analyzers illustrated in
Fig.~\ref{fig:  SeqMeas} is used to
make a joint measurement of
the corresponding target
observables, $(\mathbf{e}_1 \cdot
\hat{\mathbf{L}})^2$,
$(\mathbf{e}_2 \cdot
\hat{\mathbf{L}})^2$,
$(\mathbf{e}_3 \cdot
\hat{\mathbf{L}})^2$.  Let $\mathbf{e}'_1,
\mathbf{e}'_2, \mathbf{e}'_3
\in S'_2$ represent the actual alignments of the
analyzers.  We will assume that the precision
with which the analyzers can be aligned is very
high, so that the triad $\mathbf{e}'_1,
\mathbf{e}'_2, \mathbf{e}'_3$ is very nearly
orthonormal.  

It follows from the definitions of $f$, $\tilde{f}$
that, for each
$r$, there is probability $\ge 0.5$ that
$f(\mathbf{e}'_r)=\tilde{f}(\mathbf{e}_r)$. 
Consequently, there is a non-negligible probability
that
$\mathbf{e}'_1,
\mathbf{e}'_2, \mathbf{e}'_3$
$f$-evaluates to one of the ``illegal''
combinations $111$,
$100$, $010$, $001$, $000$ (in fact it is
straightforward, though somewhat tedious to show
that the probability of obtaining one of these
combinations is  $\ge 0.5$).
On the other hand, the fact that
the triad 
$\mathbf{e}'_1,
\mathbf{e}'_2, \mathbf{e}'_3$ is almost
orthonormal, and the results proved in 
Section~\ref{sec:  JMeasTriad}, together imply
 that the probability
 that the result of the measurement
will be one of these combinations is $\approx 0$.
It follows that there is non-negligible
probability (in fact, probability $\gtrsim 0.5$)
that, for at least one of the observables
$(\mathbf{e}'_1 \cdot
\hat{\mathbf{L}})^2$,
$(\mathbf{e}'_2 \cdot
\hat{\mathbf{L}})^2$,
$(\mathbf{e}'_3 \cdot
\hat{\mathbf{L}})^2$, the
measured value is not close to the $f$-value.

This establishes that, if the alignment errors are
random, and statistically independent, then the model
must exhibit a form of contextuality:  for it means
that the probable outcome of an approximate
measurement must, in general, be strongly dependent,
not only the observable which is being measured, but
also on the particular way in which the measurement is
carried out.  It follows that, if the stated
assumptions are true, then the model fails to satisfy
clause~\ref{en:  AOVProx} of the AOV principle (as
stated in the Introduction). 
\section{The Assumption of Independence}
\label{sec:  independence}
It is easily seen that the assumption that the
alignment errors are statistically independent is
crucial to the argument in the last section.
For instance, one can envisage a model in which the
errors are correlated in such a way that the actual
alignments are always orthogonal.  In that case the
 situation would reduce to the one considered by
MKC.  It should, however, be noted that this
assumption, besides being somewhat implausible, is
empirically falsifiable.  We have, until now, been
following MKC in assuming that the discrepancies
between target and actual alignments are a consequence
of the limited precision of the measuring device. 
However, one can equally well consider a case where 
the alignment ``errors'' are artificially controlled, 
using a random number generator, to be much
larger than the minimum attainable errors (and
consequently measurable).  The argument of the last
section applies to this case just as well as to the
case when the errors are due to the finite precision
of the instrument.  Consequently, if one wished to
avoid the conclusion to that argument in the manner
suggested, then one would have to postulate that it
is physically impossible to set up the apparatus in
such a way that the angles between the alignments are
measurably different from
$90^{\circ}$---which (quite apart from the
implausibility of the suggestion) is a definite
empirical prediction.

Of course, the fact that, in the MKC models, for each
of the values 0 and 1, the set of vectors which are
assigned that value constitute a dense subset of
$S_2$, means that these models are very flexible. 
Consequently, it may  be that
there exist other, rather more subtle, postulates
regarding the   distribution of 
actual alignments which are not empirically
falsifiable, and which do have the property that
clause~\ref{en:  AOVProx} of the AOV principle is then
satisfied.
However, it would not be very easy  to prove
that this was the case.
In a
\emph{complete} hidden variables theory, the
probability measure describing the distribution of
actual alignments is not a feature which one is free
simply to postulate.  It has to be derived, from the
dynamics of the interacting
system+apparatus+environment composite.  The models
discussed by MKC are incomplete, since they do not
include a specification of the dynamics.  It is a
highly non-trivial question, as to whether there
exists a dynamics which,  in every situation, gives
rise to a probability distribution having the desired
properties---not  only in situations where the
alignment errors arise ``naturally'', but also in
situations where the errors are adjusted ``by hand''
(in the manner described in the last paragraph).  In
the absence of a solution to this problem, the
question as to whether there exist hidden variables
theories satisfying  clause~\ref{en:  AOVProx} of the
AOV principle must be regarded as open.

Suppose, however,   that  a suitable
dynamics could be constructed.  Regarded from the
perspective of classical intuition the carefully
adjusted correlations between the different alignment
errors in a complex apparatus which such a model
must exhibit  would seem very peculiar (they
would have something of the flavour of a
``conspiracy'').  However, what is perhaps rather more
to the point is the fact that this phenomenon would
itself represent a kind of contextuality:  for it would
mean that the statistical fluctuations in an analyzer
were
\emph{ineluctably} dependent on the overall
experimental context in which the analyzer was
used.
A theory of this kind (supposing that it could be
constructed) would not so
much nullify the contextuality asserted in the
conclusion to the Kochen-Specker theorem, as change
the locus of the contextuality, from the system, onto
the alignment errors.

The  conclusion consequently seems to be that,
although one may modify its precise form, some kind of
contextuality must appear somewhere, in any hidden
variables theory.

The problem of trying to nullify the
non-classical elements in a hidden variables
theory might be compared with the problem of
trying to nullify the rucks in a badly fitted
carpet.  The carpet corresponds to the
hidden-variables theory.   The nails holding the
carpet down correspond to the empirical data. 
One has a certain amount of freedom to move the
rucks around; and with sufficient ingenuity one
may succeed in making them less noticeable. 
However, so long as the nails remain in place,
the rucks cannot actually be  eliminated.
\section{Clifton and Kent's POVM Theorem}
\label{sec:  CKonPOVMs}
In addition to the part of their argument which
we have been considering up to now, Clifton and
Kent~\cite{KentB} also  prove a theorem which,
according to them, ``rule[s] out falsifications of
non-contextual models based on generalized
observables, represented by POV measures'' (also see
Kent~\cite{KentA}).  As we have seen, an approximate
joint measurement of non-commuting observables is most
conveniently described in terms of a POVM.  Clifton
and Kent's theorem~2 also concerns measurements whose
outcome is described in terms of a POVM. 
Consequently, it  may at first sight seem that their
theorem~2 contradicts the result proved in
Section~\ref{sec:  context} of this paper.  One
purpose of this section is to show that this is not in
fact the case.  The other purpose is to point out that,
although Clifton and Kent's theorem~2 is valid if
regarded as a piece of pure mathematics, its
physical significance is more questionable.

Clifton and Kent show, that given any finite
dimensional Hilbert space $\mathcal{H}$, there exists
a set $\mathcal{A}_{\mathrm{d}}$ of positive operators
acting on
$\mathcal{H}$, and a truth function
$t_\mathcal{A} \colon
\mathcal{A}_{\mathrm{d}}\to \{0,1\}$ such that 
\begin{enumerate}
\item If $\{A_i\}\subseteq \mathcal{A}_{\mathrm{d}}$
is a finite positive operator resolution of the
identity (so that
$\sum_i A_i =1$), then
\begin{equation*}
\sum_{i} t_\mathcal{A} (A_i) = 1
\end{equation*}
\item the set of finite positive operator resolutions
of the identity contained in
$\mathcal{A}_{\mathrm{d}}$ is a countable, dense
subset of the set of all finite positive operator
resolutions of the identity (``dense''
relative to a topology defined in their paper).
\end{enumerate}
They further argue that the set of all such
functions $t_\mathcal{A}$ is sufficiently large for
the theory to be able to recover the statistical
predictions of any density operator (in so far as these
are testable using finite precision instruments).

In considering this claim we note, first of all, that
every projection is also a positive operator. 
Consequently, the set
$\mathcal{P}_{\mathrm{d}}$ of admissible projections
which features in Clifton and Kent's theorem 1 should
be contained in the set $\mathcal{A}_{\mathrm{d}}$ of
admissible positive operators which features in their
theorem 2.  Also, for any given assignment of hidden
variables, the truth function $t_{\mathcal{P}} \colon
\mathcal{P}_{\mathrm{d}} \to \{0,1\}$ should be the
restriction to $\mathcal{P}_{\mathrm{d}}$ of the
truth function $t_\mathcal{A} \colon
\mathcal{A}_{\mathrm{d}}\to \{0,1\}$.  It is not
entirely clear  from  Clifton and Kent's paper
that these requirements can be satisfied.  For the sake
of argument, we will assume that they are satisfied.

We next address the question, as to what, if any,
connection exists between Clifton and Kent's theorem~2
and the result which we proved in 
Section~\ref{sec:  context} of this paper.  It is true
that both of these results concern measurements whose
outcome is described by a POVM (positive operator
valued measure) which is not also a PVM (projection
valued measure).  However, there
is a  significant difference between the way in which
the POVM is interpreted. As we stressed in 
Section~\ref{sec:  Unsharp}, the measurements which we
consider are approximate measurements of
\emph{ordinary} observables.  The role of the POVM is
simply to provide a convenient mathematical
description of the measurement outcome.  By contrast, 
Clifton and Kent take it that the POVM's which feature
in their theorem  (and which are not also PVM's)
represent an entirely new species of
\emph{generalized} observable. 

Clifton and Kent do not provide any explicit details,
as to how these generalized observables are to be
measured.  However, they appear to assume that the
function 
$t_\mathcal{A} \colon
\mathcal{A}_{\mathrm{d}}\to \{0,1\}$ can be specified
independently of the function $t_{\mathcal{P}} \colon
\mathcal{P}_{\mathrm{d}} \to \{0,1\}$ (as we noted
above, they do not even explicitly impose the
requirement that $\mathcal{P}_{\mathrm{d}}$ should be
contained in $\mathcal{A}_{\mathrm{d}}$, and  that
$t_{\mathcal{P}}$ should be the restriction of 
$t_\mathcal{A}$).  This suggests that they are
assuming that, corresponding to the new class of 
generalized observables, there exists a new class of
generalized measurements.

In the approximate measurement procedures
which we described in Sections~\ref{sec:  MeasThy},
\ref{sec:  POVM} and~\ref{sec:  JMeasTriad} the final
result is obtained by recording the pointer
positions.  The primary mathematical construct
describing the result of the measurement is thus the
PVM which gives the distribution of pointer
positions.  The POVM is a secondary construct which is
defined in terms of this PVM [see 
Eqs.~(\ref{eq:  TDef}) and~(\ref{eq:  POVMGenCase})]. 
Let $\hat{E}_{a_1, \dots, a_n}$  be the element of the
POVM which describes the probability that the pointer
positions will be $a_1, \dots, a_n$.  Then the
admissibility of
$\hat{E}_{a_1, \dots, a_n}$, as an operator
describing the outcome of a physically possible
approximate  measurement process, entirely depends on
the admissibility of  the corresponding projection
operator, acting on the apparatus state space.  That
is, $\hat{E}_{a_1, \dots, a_n}$ is admissible if and
only if the corresponding projection belongs to the
set $\mathcal{P}_{
\mathrm{d}}^{\mathrm{ap}}$, of admissible apparatus
projections.  Moreover, if $\hat{E}_{a_1, \dots, a_n}$
is admissible, then it should be assigned the same
truth value as the corresponding apparatus
projection.  Clifton and Kent, on the other hand,
because they interpret the POVM as the mathematical
representation of a completely different kind of
observable, assume that they are free to fix the set
$\mathcal{A}_{
\mathrm{d}}$ and  truth function
$t_{\mathcal{A}}\colon \mathcal{A}_{
\mathrm{d}} \to \{0,1\}$ without having any regard for
the apparatus set
$\mathcal{P}_{
\mathrm{d}}^{\mathrm{ap}}$ and truth function
$t_{\mathcal{P}}^{\mathrm{ap}}
\colon \mathcal{P}_{\mathrm{d}}\to \{0,1\}$. 
Consequently, their theorem~2 has no bearing on the
result proved in Section~\ref{sec:  context} of this
paper.

Until now we have been concerned with the
relationship (or lack of relationship) between Clifton
and Kent's theorem 2 and the result which we proved in 
Section~\ref{sec:  context}.  However, similar
considerations also show that there are certain
obscurities regarding the significance of Clifton and
Kent's result, even when it is interpreted in their
suggested terms.  It is, of
course, the case  that not every POVM arises in the
manner discussed in this paper, as a way of describing
the outcome of an approximate measurement of an
ordinary observable.  In other contexts it may be  
appropriate to think of a POVM as representing a 
generalized observable.  But, whatever the entity
is called, one always needs to specify how it
is to be measured.  The usual answer to this question
makes use of the Neumark extension 
theorem~\cite{Peres,BuschBk,Akheiz,Holevo,PeresD}.
 In
the variant of this approach which was proposed by 
Peres~\cite{Peres,PeresD},
the system of interest is combined with an
ancilla, and an ideal  measurement is
performed on the composite.  The outcome of this
measurement is described by a PVM.  It follows that, in
this case too,  once one has fixed  the set 
$\mathcal{P}_{\mathrm{d}}$
of admissible projections for the system+ancilla
composite, one is no longer free to omit  the
corresponding system-space positive operators from the
set of admissible  positive operators.

It can be seen from this that one has  less freedom
to choose the set $\mathcal{A}_{\mathrm{d}}$ than
Clifton and Kent assume.
The situation regarding the 
truth values which should be assigned to the members of
this set is even more problematic.  It is not simply
that these values are already partially determined  by
the function
$t_{\mathcal{P}}$ describing the pointer observables
(in the case of an approximate measurement procedure
of the kind discussed in Section~\ref{sec:  POVM}), or
the system+ancilla composite (in the case of Peres'
version of the Neumark construction).  It is not even
clear that the truth values are determined
unambiguously; for it may happen that  a given POVM
can be physically realized in more than one way. 
Suppose, for example, that the POVM described in
Section~\ref{sec:  JMeasTriad} is alternatively
realized by Peres' version of the Neumark
construction.  In both cases, the outcome is fixed
once the relevant functions
$t_{\mathcal{P}}$ are fixed; but it is far from
clear that the outcomes will be the same.  In short,
it is questionable whether it is appropriate to think
in terms of there being a well-defined truth function
on the  set
$\mathcal{A}_{\mathrm{d}}$. 
\section{Conclusion}
In this paper we have made a number of criticisms of
the arguments of MKC.  This should not be allowed to
obscure the fact that their work is, in our view, both
interesting and  valuable.  MKC have taken a
question which seemed clear-cut, and shown that it is
in fact much more subtle and intricate than had
previously been appreciated.  They have thereby
significantly deepened our understanding of the
conceptual implications of quantum mechanics.

Similar qualifications apply to our critical remarks
(based on Uffink's~\cite{Uffink} criticisms)
concerning the use of the concept of an unsharp
observable to describe approximate measurements of
ordinary observables.  As we stressed at the end of  
Section~\ref{sec:  Unsharp}, these criticisms should
not be taken to imply that we question the need for an
extended concept of physical observable.


\begin{thebibliography}{99}
\label{sec:  bibliography}
\bibitem{Meyer}
D.A.~Meyer,
\emph{Phys.\ Rev.\ Lett.}\ \textbf{83}, 3751
(1999).
\bibitem{KentA}
A.~Kent,
\emph{Phys.\ Rev.\ Lett.}\ \textbf{83}, 3755 
(1999).
\bibitem{KentB}
R.~Clifton and A.~Kent,
\emph{Proc.\ Roy.\ Soc.\ A}, to appear.
Also available as Los Alamos e-print
quant-ph/9908031.
\bibitem{Koch}
S.~Kochen and E.P.~Specker, 
\emph{J.\ Math.\ Mech.}\ \textbf{17}, 59 (1967).
\bibitem{Bell}
J.S.~Bell,
\emph{Rev.\ Mod.\ Phys.}\ \textbf{38}, 447 (1966).
\bibitem{MerminB}
N.D.~Mermin,
\emph{Rev.\ Mod.\ Phys.}\ \textbf{65}, 803 (1993).
\bibitem{Peres}
A.~Peres,
\emph{Quantum Theory:  Concepts and Methods}
(Kluwer, Dordrecht, 1993).
\bibitem{Mermin}
N.D.~Mermin,
Los Alamos e-print, quant-ph/9912081.
\bibitem{Havli}
H.~Havlicek, G.~Krenn, J.~Summhammer and
K.~Svozil, Los Alamos e-print quant-ph/9911040.
\bibitem{Cabel}
A.~Cabello,
Los Alamos e-print quant-ph/9911024.
\bibitem{Basu}
S.~Basu, S.~Bandyopadhyay, G.~Kar and D.~Home,
Los Alamos e-print, quant-ph/9907030.
\bibitem{self1a}
D.M.~Appleby,
\emph{Int.\ J.\ Theor.\ Phys.}\ \textbf{37}, 1491
(1998).
\bibitem{self1b}
D.M.~Appleby,
\emph{Int.\ J.\ Theor.\ Phys.}\ \textbf{37}, 2557
(1998).
\bibitem{Witt}
L.~Wittgenstein,
\emph{Philosophical Investigations}
(Basil Blackwell, Oxford, 1967).
\bibitem{PosMom}
E.~Arthurs and S.C.~Kelly,
\emph{Bell Syst.\ Tech.\ J.}\ \textbf{44}, 725
(1965);
P.~Busch, 
\emph{Int.\ J.\ Theor.\ Phys.}\ \textbf{24}, 63
(1985);
S.L.~Braunstein, C.M.~Caves and
G.J.~Milburn,
\emph{Phys.\ Rev.\ A} \textbf{43}, 1153 (1991);
S.~Stenholm,
\emph{Ann.\ Phys.}\ \textbf{218}, 233 (1992);
U.~Leonhardt and H.~Paul,
\emph{J.\ Mod.\ Opt.}\ \textbf{40}, 1745 (1993);
M.G.~Raymer, \emph{Am.\ J.\ Phys.}\ \textbf{62},
986 (1994);
U.~Leonhardt, B.~B\"{o}hmer and
H.~Paul,
\emph{Optics Commun.}\ \textbf{119}, 296 (1995);
P.~T\"{o}rma, S.~Stenholm and I.~Jex,
\emph{Phys.\ Rev.\ A} \textbf{52}, 4812 (1995);
W.L.~Power, S.M.~Tan and M.~Wilkens,
\emph{J.\ Mod.\ Opt.}\ \textbf{44}, 2591 (1997).
\bibitem{self1c}
D.M.~Appleby,
\emph{J.\ Phys.\ A} \textbf{31}, 6419 (1998).
\bibitem{self2}
D.M.~Appleby,
\emph{Int.\ J.\ Theor.\ Phys.}\ \textbf{38}, 807
(1999);
\emph{J.\ Mod.\ Opt.}\ \textbf{46}, 813 (1999).
\bibitem{spin}
E.~Prugovecki, 
\emph{J.\ Phys.\ A},
\textbf{10}, 543 (1977);
F.E.~Schroeck,
\emph{Found.\ Phys.}\ \textbf{12}, 479 (1982);
P.~Busch, \emph{Phys.\ Rev.\ D} \textbf{33}, 2253
(1986); 
P.~Busch,
\emph{Found.\ Phys.}\ \textbf{17}, 905 (1987);
P.~Busch,
\emph{Phys.\ Lett.\ A} \textbf{130}, 323 (1988);
P.~Busch and F.E.~Schroeck,
\emph{Found.\ Phys.}\ \textbf{19}, 807 (1989);
M.~Grabowski,
\emph{Int.\ J.\ Theor.\ Phys.}\ \textbf{28}, 1215
(1989);
 H.~Martens and W.M.~de Muynck,
\emph{J.\ Phys. A} \textbf{26}, 2001 (1993);
P.~Kienzler,
\emph{Int.\ J.\ Theor.\ Phys.}\ \textbf{37}, 257
(1998).
\bibitem{self3}
D.M.~Appleby,
Los Alamos e-print, quant-ph/9911021.
\bibitem{BuschBk}
P.~Busch, M.~Grabowski and P.J.~Lahti,
\emph{Operational Quantum Physics}
(Springer-Verlag, Berlin,1995).
\bibitem{LeonBk}
U.~Leonhardt,
\emph{Measuring the Quantum State of Light}
(Cambridge University Press, Cambridge, 1997).
\bibitem{Uffink}
J.~Uffink,
\emph{Int.\ J.\ Theor.\ Phys.}\ \textbf{33},
199 (1994).
\bibitem{Flem}
G.N.~Fleming,
\emph{Stud.\ Hist.\ Phil.\ Mod.\ Phys.}\
\textbf{31}, 117 (2000).
\bibitem{PeresB}
A.~Peres,
\emph{Phys.\ Rev.\ A} \textbf{61}, 022116 (2000).
\bibitem{Dirac}
P.A.M.~Dirac,
\emph{The Principles of Quantum Mechanics}
(Clarendon Press, Oxford, 1958).
\bibitem{VNeu}
J.~von Neumann,
\emph{Mathematische Grundlagen der Quantenmechanik}
(Springer, Berlin, 1932); [English translation: 
\emph{Mathematical Foundations of Quantum Mechanics}
(Princeton University Press, Princeton NJ, 1955)].
\bibitem{BuschB}
P.~Busch, M.~Grabowski and P.J.~Lahti,
\emph{Ann.\ Phys.\ (NY)} \textbf{237}, 1 (1995).
\bibitem{Pegg}
D.T.~Pegg and S.M.~Barnett,
\emph{J.\ Mod.\ Opt.}\ \textbf{44}, 225 (1997).
\bibitem{Buzek}
V.~Bu\v{z}ek, R.~Derka and S.~Massar,
\emph{Phys.\ Rev.\ Lett.}\ \textbf{82}, 2207 (1999).
\bibitem{Opp}
J.~Oppenheim, B.~Reznik and W.G.~Unruh,
\emph{Phys.\ Rev.\ A} \textbf{59}, 1804 (1999).
\bibitem{Muga}
I.L.~Egusquiza and J.G.~Muga,
\emph{Phys.\ Rev.\ A} \textbf{61}, 012104 (2000).
\bibitem{PeresC}
A.~Peres,
\emph{J.\ Phys.\ A} \textbf{24}, L175 (1991).
\bibitem{Akheiz}
N.I.~Akhiezer and I.M.~Glazman,
\emph{Theory of Linear Operators in Hilbert Space},
vol.~2 (Frederick Ungar, New York, 1963).
\bibitem{Holevo}
A.S.~Holevo,
\emph{Probabilistic and Statistical Aspects of Quantum
Theory} (North Holland, Amsterdam, 1982).
\bibitem{PeresD}
A.~Peres,
\emph{Found.\ Phys.}\ \textbf{20}, 1441 (1990).
\end{thebibliography}
\end{document}